\documentclass[12pt,preprint]{aastex}

\slugcomment{Draft version November 13, 2007.}

\shorttitle{Fluorine in RCBs}
\shortauthors{Pandey et al.}

\begin{document}

\title{Fluorine in R Coronae Borealis Stars}

\author{Gajendra Pandey}
\affil{Indian Institute of Astrophysics;
Bangalore,  560034 India}
\email{pandey@iiap.res.in}

\author{David L.\ Lambert}
\affil{The W.J. McDonald Observatory, University of Texas at Austin; Austin, 
TX 78712-1083}
\email{dll@astro.as.utexas.edu}

\and

\author{N. Kameswara Rao}
\affil{Indian Institute of Astrophysics;
Bangalore,  560034 India}
\email{nkrao@iiap.res.in}

\begin{abstract}
Neutral fluorine (F\,{\sc i}) lines are identified in the optical
spectra of several R\,Coronae Borealis stars (RCBs) at maximum light.
These lines provide the first measurement of the fluorine
abundance in these stars. Fluorine is enriched in some RCBs by factors of 800 to 8000
relative to its likely initial abundance.
The overabundances of fluorine are evidence for the synthesis 
of fluorine. These results are discussed in the light of the scenario that
RCBs are formed by accretion of an He white dwarf by a C-O white dwarf.
Sakurai's object (V4334\,Sgr), a final He-shell flash product, shows no
detectable F\,{\sc i} lines.
\end{abstract}

%\clearpage

\section{Introduction}

R Coronae Borealis (RCB) stars comprise a sequence of hydrogen-deficient
supergiants with effective temperatures from about $T_{\rm eff} = 3500$ K,
as represented by Z\,UMi and DY\,Per, to about 19,500 K, as represented
by DY\,Cen. 
The characteristic of H-deficiency is shared by the H-deficient cool
carbon (HdC) stars at low temperatures and by the extreme helium
(EHe) stars at high temperatures. A common assumption is that the
sequence HdC - RCB - EHe in the ($T_{\rm eff},\log g$) plane
reflects a close evolutionary connection. In this sequence, the
RCBs are distinguished from HdC and EHe stars by their second principal defining
characteristic: their propensity to fade, in visual light, unpredictably as a
cloud of carbon dust obscures the star. This propensity is not universally shared: 
XX\,Cam has yet to be observed below maximum light. In addition, DY\,Cen might be
considered as an EHe star known to experience RCB-like fadings.

If HdC, RCB, and EHe stars share a common heritage, the expectation
is that their atmospheric compositions should show some common
features \citep{pan2004a,rao2005}. It is through the compositions that one hopes to test
theoretical ideas about the origins of these extremely rare
stars; just five HdC, about 40 RCB \citep{zanie2005}, and 21 EHe 
stars are known in the Galaxy. Currently, two
scenarios remain in contention to account for these
H-poor high luminosity stars. In the first, a final He-shell flash in a
post-AGB star on the white dwarf cooling track creates a H-poor luminous
star. This is dubbed the `final flash' (FF) scenario. In the second, the H-poor
star is formed from a merger of a He white dwarf with a C-O white dwarf.
In a close binary system, accretion of the He white dwarf by the C-O white dwarf
may lead to a H-poor supergiant with the C-O white dwarf as its core. This is
called the `double degenerate' (DD) scenario.

Products of both the FF and DD scenarios may be presumed to exist.
A determination of which scenario provided which star rests in large
part on the observed chemical composition of a star's atmosphere and theoretical
predictions about the FF and DD products. Evidence from
elemental abundances, especially the H, C, N, and O abundances, suggests that 
the RCB and EHe stars evolved via the
DD rather than the FF route \citep{asp00,pan2001,saio02,pan2006}.
Convincing, essentially incontrovertible, evidence
that the DD scenario led to the HdCs and some cool RCBs was
presented by \citet{clay2007} with their discovery that
the $^{18}$O was very abundant in their atmospheres. This usually
rare isotope of oxygen was attributed to nucleosynthesis occurring
during and following accretion of the He-rich material onto the C-O white dwarf.

Determination of the oxygen isotopic ratios demands
a cool star with the CO vibration-rotation bands in its spectrum.
The majority of RCBs and all of the EHes are too hot for CO to contribute 
to their spectra \citep{tenen2005}.\footnote{A possibility exists that 
circumstellar CO may be detectable for the RCBs. Cool gas containing 
C$_2$ molecules has been seen for V854\,Cen, R\,CrB, and V\,CrA when the 
stars were in decline \citep{raolamb2000,raolambsh2006,raolamb2007}.}
An alternative tracer of nucleosynthesis during a merger may
be provided by the fluorine abundance. Considerable enrichment
of EHe stars with F was discovered by \citet{pana2006} from detection and
analysis of about a dozen F\,{\sc i} lines in optical spectra of
cool EHe stars. Clayton et al.'s (2007) calculations suggest that F synthesis is
possible in the DD scenario. In this paper, we report on a search for
F\,{\sc i} lines in spectra of RCBs and discuss the F abundances in
light of the results for EHes and the expectations for the DD and FF
scenarios.

\section{Observations}

High-resolution optical spectra of RCBs at maximum light obtained
at the W.J. McDonald Observatory and the Vainu Bappu Observatory
were examined for the presence of  F\,{\sc i} lines.  A key requirement 
of the spectra for inclusion in the study was that the absorption
line profiles be symmetric; an asymmetry and a doubling of the lines
is attributable to a pulsation of the atmosphere that may induce a
shock wave. An additional requirement was that the S/N ratio be
adequate for detection of weak lines.

Spectra from the
McDonald Observatory were obtained with the 2.7 meter Harlan J.
Smith telescope and the coud\'{e} cross-dispersed
echelle spectrometer \citep{tull95} at a resolving power 
$R = \lambda/\Delta\lambda = 60,000$ with a
bandpass covering much of the 3800 -- 10000 \AA\ interval.

At the Vainu Bappu Observatory (VBO), spectra were acquired with the Vainu Bappu
Telescope with a fiber-fed cross-dispersed echelle spectrometer \citep{rao04,rao05b}.
These spectra were acquired with two different grating 
settings to cover the wavelength regions of the crucial F\,{\sc i} lines and
are at a resolving power $R \simeq 30,000$ 
with a wavelength coverage of 4860\AA\ to 7900\AA.

All spectra were reduced in standard fashion using the Image Reduction
and Analysis Facility (IRAF) software.\footnote{The IRAF software is 
distributed by the National Optical Astronomy Observatories under contract 
with the National Science Foundation}
The reduction procedures usually included the removal of
the telluric absorption lines through an interactive process (task
{\it telluric}) using a spectrum of an early-type rapidly rotating
star.

Suitable McDonald spectra of 13 RCBs were available for analysis. In
several cases, spectra of the same star from different epochs were
available. The stars are listed in Table 1. 
%{\bf This can be the table in
%which we list stellar parameters and the F abundances.}
At VBO, spectra were obtained of UW\,Cen, a star inaccessible from
the McDonald Observatory, and also of XX\,Cam, R\,CrB, RY\,Sgr, and VZ\,Sgr.
The UW\,Cen spectrum was complemented by  a CTIO spectrum from 1992 (see
Asplund et al. 2000).

\begin{deluxetable}{llcccccr}
\tabletypesize{\scriptsize}
\tablewidth{0pt}
\tablecolumns{7}
%\tablewidth{0pc}
\setcounter{table} {0}
\tablecaption{The analyzed RCBs, their stellar parameters,
and fluorine abundances from individual F\,{\sc i} lines. Sakurai's object, a final He-shell flash product, is also listed.}
\tablehead{
\colhead{Star} & \colhead{($T_{\rm eff}$, $\log g$, $\xi$)} & \multicolumn{6}{c}{log $\epsilon(\rm F)$}\\
\cline{3-8} \\
\colhead{} & \colhead{} & \colhead{7398.68\AA} & \colhead{7754.69\AA} & \colhead{6902.47\AA} & \colhead{7425.64\AA} & \colhead{6834.26\AA} & \colhead{Mean}}
%\colhead{} & \colhead{} & \colhead{}}
\startdata
V3795\,Sgr$^a$ & (8000, 1.0, 10.0) & \nodata & 6.60 & 6.65 & 6.70 & 6.70 & 6.66$\pm$0.05(4)\\
&&&&&&&\\
UW\,Cen & (7500, 1.0, 12.0) & 7.20 & \nodata & 7.00 & 7.10 & 7.20 & 7.1$\pm$0.1(4)\\
&&&&&&&\\
RY\,Sgr & (7250, 0.75, 6.0) & \nodata & \nodata & 6.80 & \nodata & 7.10 & 6.95$\pm$0.2(2) \\
&&&&&&&\\
XX\,Cam & (7250, 0.75, 9.0) & \nodata & \nodata & $<$5.6 & \nodata & $<$5.6 & $<$5.6 \\
&&&&&&&\\
UV\,Cas & (7250, 0.5, 7.0) & \nodata & \nodata & 6.20 & \nodata & \nodata & 6.2(1)\\
&&&&&&&\\
UX\,Ant & (7000, 0.5, 5.0) & \nodata & \nodata & $<$6.2 & \nodata & $<$6.2 & $<$6.2\\
&&&&&&&\\
VZ\,Sgr$^a$ & (7000, 0.5, 8.0) & \nodata & \nodata & 6.30 & \nodata & 6.50 & 6.4$\pm$0.1(2)\\
&&&&&&&\\
R\,CrB & (6750, 0.5, 7.0) & \nodata & \nodata & 6.85 & \nodata & 7.00 & 6.9$\pm$0.1(2)\\
&&&&&&&\\
V2552\,Oph & (6750, 0.5, 7.0) & \nodata & \nodata & 6.60 & \nodata & \nodata & 6.6(1)\\
&&&&&&&\\
V854\,Cen$^a$ & (6750, 0.0, 6.0) & \nodata & \nodata & \nodata & \nodata & $<$5.7 & $<$5.7\\
&&&&&&&\\
SU\,Tau & (6500, 0.5, 7.0) & \nodata & \nodata & 6.90 & \nodata & 7.00 & 6.95$\pm$0.1(2)\\
&&&&&&&\\
V\,CrA$^a$ & (6500, 0.5, 7.0) & \nodata & \nodata & 6.5: & \nodata & \nodata & 6.5:(1)\\
&&&&&&&\\
V482\,Cyg & (6500, 0.5, 4.0) & \nodata & \nodata & \nodata & \nodata & 6.6: & 6.6:(1)\\
&&&&&&&\\
GU\,Sgr & (6250, 0.5, 7.0) & \nodata & \nodata & \nodata & \nodata & 7.2: & 7.2:(1)\\
&&&&&&&\\
FH\,Sct & (6250, 0.25, 6.0) & \nodata & \nodata & \nodata & \nodata & 7.2: & 7.2:(1)\\
&&&&&&&\\
Sakurai's object & (7500, 0.0, 8.0) & \nodata & \nodata & \nodata & \nodata & $<$5.4 & $<$5.4\\
&&&&&&&\\
\enddata
\tablenotetext{a}{Minority RCBs}
%\tablenotetext{b}{See text}
\end{deluxetable}

%{\bf I'd suggest that we introduce Sakurai's object at an appropriate
%point in the discussion and not mention it in this section.}

\clearpage

\begin{figure}
\epsscale{1.00}
\plotone{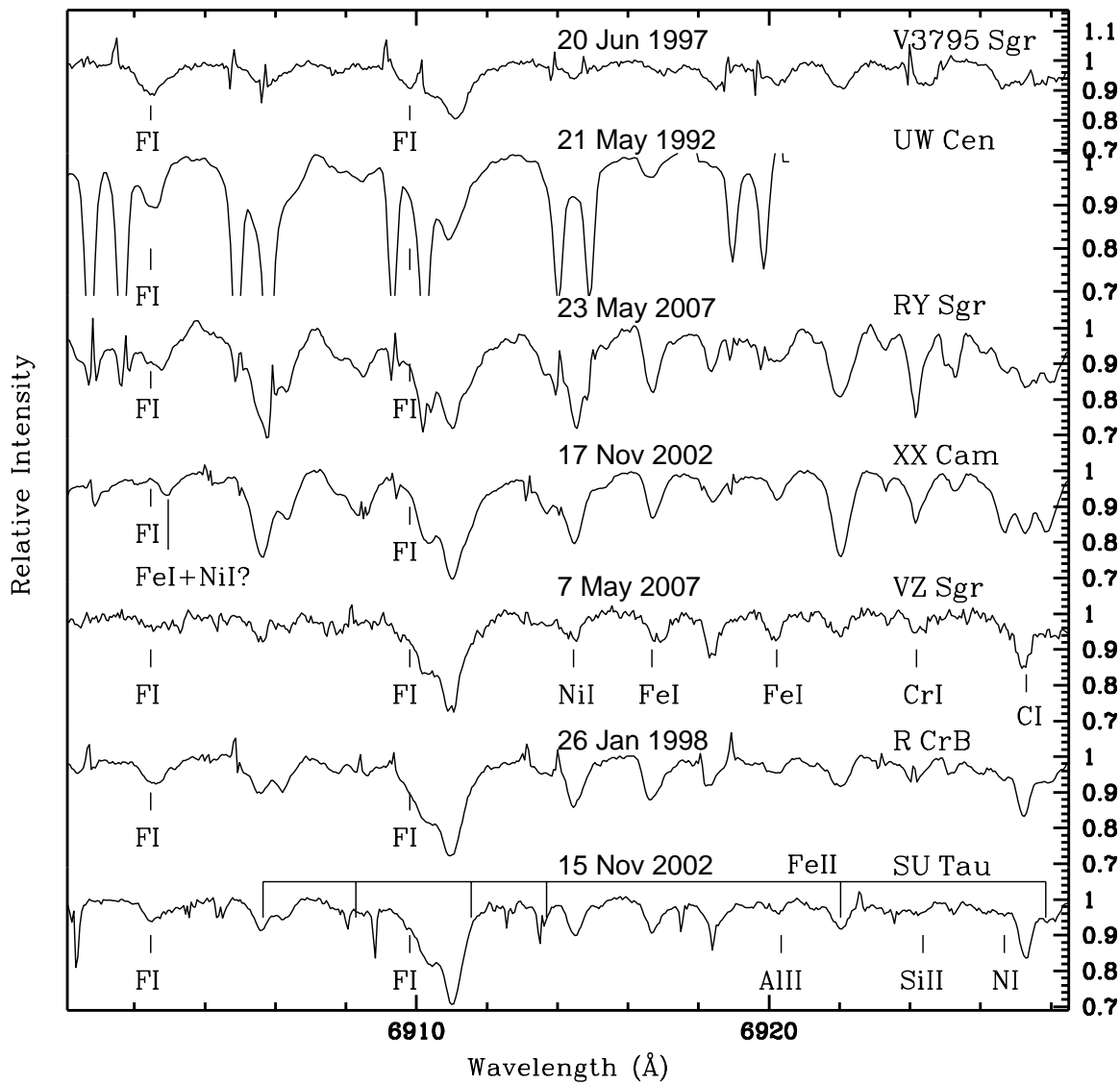}
\caption{The spectral region from 6900 -- 6928 \AA\ is shown for seven RCBs with the
hottest star at the top and the coolest star at the bottom. The F\,{\sc i}
line at 6902.47 \AA\ labelled below each spectrum is obviously
present in V3795\,Sgr and UW\,Cen  (the hottest two stars), and absent from
XX\,Cam but a likely contributor to the spectra of the other stars. A
weaker F\,{\sc i} line at 6909.81 \AA\ is also labelled but is either
lost in the wing of a stronger line and/or masked by a telluric O$_2$
line. Some other atomic lines are identified that apply to all spectra. \label{fig1}}
\end{figure}

\begin{figure}
\epsscale{1.00}
\plotone{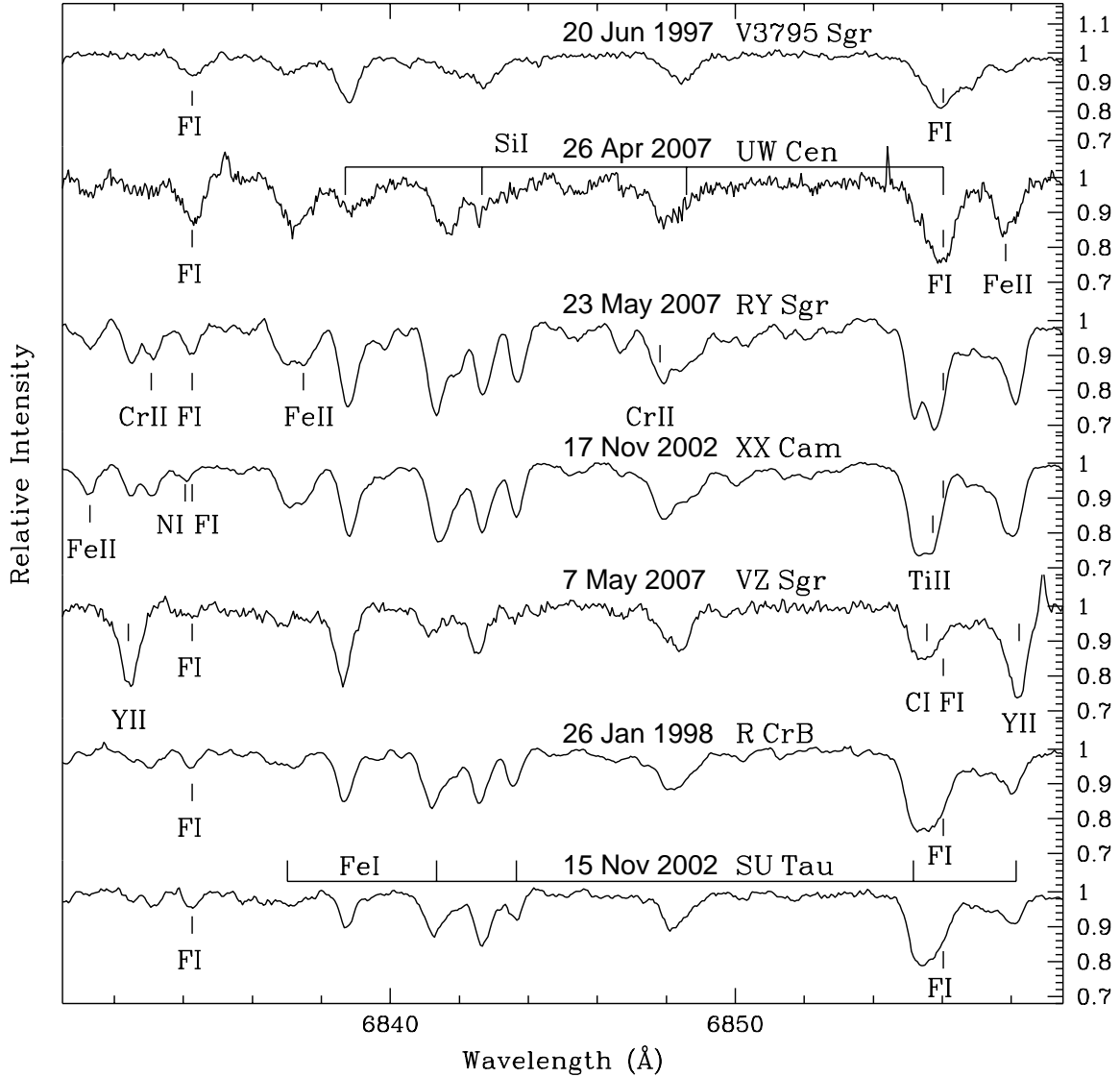}
\caption{The spectral region from 6830 -- 6860 \AA\ is shown for seven RCBs with the
hottest star at the top and the coolest star at the bottom. The F\,{\sc i}
line at 6834.26 \AA\ labelled below each spectrum is obviously
present in V3795\,Sgr and UW\,Cen  (the hottest two stars), and absent from
XX\,Cam but a likely contributor to the spectra of the other stars. A  stronger
F\,{\sc i} line at 6856.03 \AA\ is also labeled.
This line is clearly present for V3795\,Sgr and UW\,Cen. In cooler
stars, the F\,{\sc i} contribution is blended severely with a Si\,{\sc i}
and a Fe\,{\sc i} line.
Some other atomic lines are identified that apply to all spectra. \label{fig2}}
\end{figure}

\clearpage

\section{F\,{\sc i} Lines}

In the atmosphere of an RCB star, fluorine is present as neutral
atoms. The leading lines of F\,{\sc i} in optical spectra come from excited levels
with lower excitation potentials of 12.7 eV or higher. The
$2p^4(^3P)3s$ -- $2p^4(^3P)3p$ transition array provides 27 lines from
six multiplets with lower excitation potentials 12.70 -- 13.02 eV.
The adopted $gf$-values come from \citet{musie99}
who measured relative
transition probabilities for lines in the $3s-3p$ (and $3p-3d$) array
using emission lines from a wall-stabilized arc and used Opacity Project
calculations of selected strong lines to place results on an absolute
scale. The $gf$-values are estimated to be accurate to 15\% to 20\%.

The predicted equivalent widths ($W_\lambda$) of F\,{\sc i} in RCB stars are
sensitive to the effective temperature ($T_{\rm eff}$).
This sensitivity is highlighted with a simple calculation of the
predicted  $W_\lambda$ for the 
F\,{\sc i} 6902.47\AA\ line, the fourth strongest line in the transition
array,  with a $gf$-value of 1.51 for models with
a number density ratio C/He =1\%. The $W_\lambda$  varies from 75 m\AA\ at
$(T_{\rm eff},\log g$) = (8000 K, 1.0) to 35 m\AA\ at
(7000 K, 0.5) to 12 m\AA\ at (6250 K, 0.5) where the
($T_{\rm eff},\log g$) combinations are representative of V3795\,Sgr,
VZ\,Sgr, and GU\,Sgr, respectively, and the F abundance chosen is representative 
of values reported for EHe stars \citep{pana2006}.
These predictions suggest that opportunities for detection of
F\,{\sc i} lines will be best in the hottest RCB stars.
The high $T_{\rm eff}$ sensitivity, perhaps suggested by the excitation
potential of 13 eV, is moderated by the excitation potential (about
9 -- 11 eV) of the high levels of the carbon atom providing the continuous
opacity by photoionization.
Note that the sensitivity of predicted $W_\lambda$ to $\log g$
at a given $T_{\rm eff}$ is not appreciable.
The calculations also show that the weakest lines in the transition array (say,
$gf$ $< 0.3$) will not be detectable in cooler RCBs.

The following F\,{\sc i} lines were  identified as the principal or
leading contributor to a stellar line: 7398.68, 7754.69, 6902.47, 7425.65, 
and 6834.26\AA.
Figures 1 and 2 illustrate the windows that represent a couple of these F\,{\sc i} 
lines, where the RCBs are ordered from top to bottom in order
of decreasing effective temperature.
All spectra are aligned to the rest wavelengths of  well known lines that fall in
the wavelength regions. 

For all F\,{\sc i} lines, a search was made for blending lines.
Databases examined included the Kurucz database\footnote{http://kurucz.harvard.edu},
tables of spectra of H, C, N, and O \citep{moor93} and
Si \citep{moore65,moore67}, Vienna Atomic Line Database (VALD), and
the new Fe\,{\sc i} multiplet table \citep{nave94}.
The following remarks describe the quintet of useful lines in order to decreasing
$gf$-value:

{\bf 7398.68\AA} ($\log gf = +0.24$). This line is in
an inter-order gap for  most  McDonald spectra.
The exceptions are  McDonald spectra of XX\,Cam, UV\,Cas, 
and SU\,Tau and the VBT spectrum of UW\,Cen. The line is blended with
the  N\,{\sc i} 7398.63\AA\ line in XX\,Cam,
UV\,Cas, and SU\,Tau but the F\,{\sc i} line's contribution is obtainable
for  UW\,Cen. 
The CN (4,1) 7398.473\AA\ may be contributing to the blue wing of this line for 
cooler RCBs.
% there is a suggestion of this CN line in SU\,Tau's spectrum.

{\bf 7754.69\AA} ($\log gf = +0.24$). The major contributors to the red wing of this line are
Sm\,{\sc ii} 7755.072\AA, Mg\,{\sc i} 7755.236\AA, and Fe\,{\sc ii} 7755.563\AA.
CN(5,2) 7753.767\AA\ may be contributing to the blue wing of this line for cooler RCBs.
The F\,{\sc i} line's contribution for V3795\,Sgr is measureable.

{\bf 6902.47\AA} ($\log  gf = +0.18$). This region is crossed by telluric absorption (O$_2$) lines.
From the available database of RCB spectra at maximum light, only those spectra were 
used in which the 6902.47\AA\ stellar feature avoids these 
telluric absorption lines (see Figure 1). Also, further care has been taken to remove these
telluric absorptions by ratioing these spectra with a spectrum of an
early-type rapidly rotating star. 
This line  is present in V3795\,Sgr and UW\,Cen with a depth
comparable to that of the 7754.69\AA\ line of similar $gf$-value.
There is obviously a line present to the red that dominates in XX\,Cam and other
Fe-rich RCBs. The suspected lines are Ni\,{\sc i} at 6902.791\AA\ (VALD), and 
Fe\,{\sc i} at 6902.956\AA\ (Kurucz's database).
Through spectrum synthesis (see Section 4 for details), the line provides an estimate of or an upper limit
to the F abundance.

For the spectrum synthesis of the  region,
we take Fe\,{\sc i}  to be the significant blend 
in the RCB stars' spectra instead of the Ni\,{\sc i} line. 
XX\,Cam's spectra clearly shows this Fe\,{\sc i} 6902.956\AA\ line with little 
or no contribution from the F\,{\sc i} 6902.47\AA\ line (Figure 1).
 We estimate the $gf$-value 
of this Fe\,{\sc i} 6902.956\AA\ line by inversion using the XX\,Cam's spectrum; the
Fe abundance is taken from \citet{asp00}. Figure 3 shows the spectrum synthesis 
of this F\,{\sc i} line region for several RCBs.

{\bf 7425.64\AA} ($\log gf = -0.19$) This line is detected in V3795\,Sgr and UW\,Cen but not 
in other RCBs for which the upper limit to the F abundance is consistent with
estimates from other lines.

{\bf 6834.26\AA} ($\log gf = -0.21$). This line contributes in V3795\,Sgr and UW\,Cen, and shows its
presence in other RCBs (Figures 2  and 4).
 This line is used for
abundance determination after accounting for all the known blends. 
 Blends include La\,{\sc ii} 6834.009\AA, 
CN(7,3) 6834.236\AA\ and
N\,{\sc i} 6834.074\AA.
Kurucz's database also lists a La\,{\sc ii} 6837.904\AA\ line which is about
the same strength as the 
 La\,{\sc ii} 6834.009\AA\ line. However, the 
La\,{\sc ii} 6837.904\AA\ line is not detected even in the $s$-process enriched
RCBs.
For the N\,{\sc i} line,
we take the $gf$-value  from
Kurucz's database. XX\,Cam's spectrum essentially has little or no contribution
from the F\,{\sc i} 6834.26\AA\ line (Figure 2). The predicted N\,{\sc i} profile at 6834.074\AA\
reproduces the observed profile for XX\,Cam; N abundance is from \citet{asp00}.
Also, the equivalent widths of the common N\,{\sc i} lines measured in our RCB stars'
spectra are in agreement with the \citet{asp00} measurements within the
expected uncertainties. We adopt the N abundances from \citet{asp00} for
the analyzed RCBs except for SU\,Tau's N abundance that is from \citet{pan2004b}.
Figure 4 shows the spectrum synthesis of this F\,{\sc i} line region for several RCBs.

Where they contribute, CN Red System lines are included with their $gf$-values
taken from Kurucz's database. 
Note that  CN lines listed in Kurucz's database with 
predicted or extrapolated energies are not included in the 
spectrum synthesis. The CN strengths are determined from the 
strongest CN lines observed across the wavelength region of the
analyzed RCBs. CN contributes to these F\,{\sc i} lines only in the cooler members of 
the RCBs. Note that the CN (2,0) and CN (3,0) bandheads are clearly present in the 
cooler members of the RCBs; CN bands are typically strong in visible spectra
of cool RCB stars (e.g., Clayton 1996 and references therein). 

A search was made for all other lines in the transition array but with the
exception of clear indications of a F\,{\sc i} contribution for
V3795\,Sgr and UW\,Cen, the F\,{\sc i} line is masked by blending lines in the
other RCBs. 

\section{Abundance analysis}

The fluorine abundances were determined from the
F\,{\sc i} lines  by adopting the  procedure described
in \citet{pan2006}.
Uppsala H-deficient model atmospheres \citep{asp97a} are used with the Uppsala spectral synthesis LTE code BSYNRUN
to compute either the equivalent width of a F\,{\sc i} line or
a synthetic spectrum for a selected spectral window. The synthetic spectrum was
convolved with a Gaussian profile to account for the combined effects of the stellar
macroturbulence, rotational broadening, and the instrumental profile, before matching 
with the observed spectrum. The fact that several F\,{\sc i} lines are blended with other lines
requires that the fluorine abundance be extracted by spectrum synthesis.
Although F\,{\sc i} 6902.47\AA\ and F\,{\sc i} 6834.26\AA\ lines are blended
in most of the analyzed RCBs, we could
account for the contributing blending lines and extract the F\,{\sc i}
component just from these lines.
For the two hottest RCBs, fluorine was also measured from the F\,{\sc i} 7425.64\AA\
and F\,{\sc i} 7398.68\AA\ lines in UW\,Cen, and in V3795\,Sgr from 
the F\,{\sc i} 7425.64\AA\ and F\,{\sc i} 7754.69\AA\ lines.
The number of usable F\,{\sc i} lines declines from four for V3795\,Sgr and
UW\,Cen, the hottest RCBs, to two or zero in the coolest RCBs. This
trend reflects the decline of the predicted equivalent width for constant F
abundance with decreasing $T_{\rm eff}$. As a reminder, we note that
11 to 17 F\,{\sc i} lines were usable for the EHes with $T_{\rm eff}$'s
from 8750 K to 12750 K.

%\clearpage
%\begin{center}\small{Table 3} \\
%Fluorine abundances from individual F\,{\sc i} lines for
%the analyzed RCBs \\
%\begin{tabular}{lccccc}
%\hline
%\hline
% &  &  & log $\epsilon$(F) & & \\
%\cline{2-6}
% & 6902.47\AA\ & 6834.26 \AA\ & 7754.69\AA\ & 7127.89\AA\ & 7425.64\AA\ \\
%\hline
%V3795\,Sgr & 6.65 & 6.70 & 6.60 & $<$6.94 & $<$6.80\\
%&&&&&\\
%UW\,Cen    & 7.00 & 7.20 & 7.45 & \nodata & \nodata \\
%&&\\
%RY\,Sgr    & 6.80 & 7.10 & 7.00 & \nodata & $<$7.50 \\
%&&\\
%XX\,Cam    & $<$5.50 & $<$5.50 & 6.30 & \nodata & $<$5.60 \\
%&&\\
%UV\,Cas    & 6.20   & 6.20   & 6.50 & \nodata & $<$5.60 \\
%&&\\
%UX\,Ant    & $<$6.20 & $<$6.20 & $<$6.70 & \nodata & $<$6.80\\
%&&\\
%VZ\,Sgr    & 6.30    & 6.50    & 6.50   & $<$7.10 & $<$6.10\\
%&&\\
%R\,CrB     & 6.85    & 7.00    & 7.05   & \nodata & $<$7.20\\
%&&\\
%V2552\,Oph & 6.60    & 6.70    & 6.90   & \nodata & $<$7.10 \\
%&&\\
%V854\,Cen  & \nodata & $<$5.70 & 6.30   & \nodata & $<$5.70 \\
%&&\\
%SU\,Tau    & 6.90    & 7.00    & 7.40   & \nodata & $<$7.30 \\
%&&\\
%V\,CrA     & 6.50:   & 6.90:   & \nodata & \nodata & \nodata \\
%&&\\
%V482\,Cyg  & 6.70:   & 6.60:   & 7.00:   & \nodata & \nodata \\
%&&\\
%GU\,Sgr    & 6.90:   & 7.20:   & 6.70:   & \nodata & \nodata \\
%&&\\
%FH\,Sct    & 6.90:   & 7.20:   & 6.90:   & \nodata & \nodata \\
%&&\\
%Sakurai's object &  $<$4.80   & $<$5.20 & \nodata & \nodata & \nodata \\
%\hline
%\end{tabular}
%\end{center}
%%$^a$  This paper for the model ($T_{\rm eff}$,$\log g, \xi$) $\equiv$
%% (18300, 2.2, 16.0)\\
%%$^b$ Jeffery (1993) for the mo (18300, 2.2, 16.0)\\
%\clearpage

\clearpage

\begin{figure}
\epsscale{1.00}
\plotone{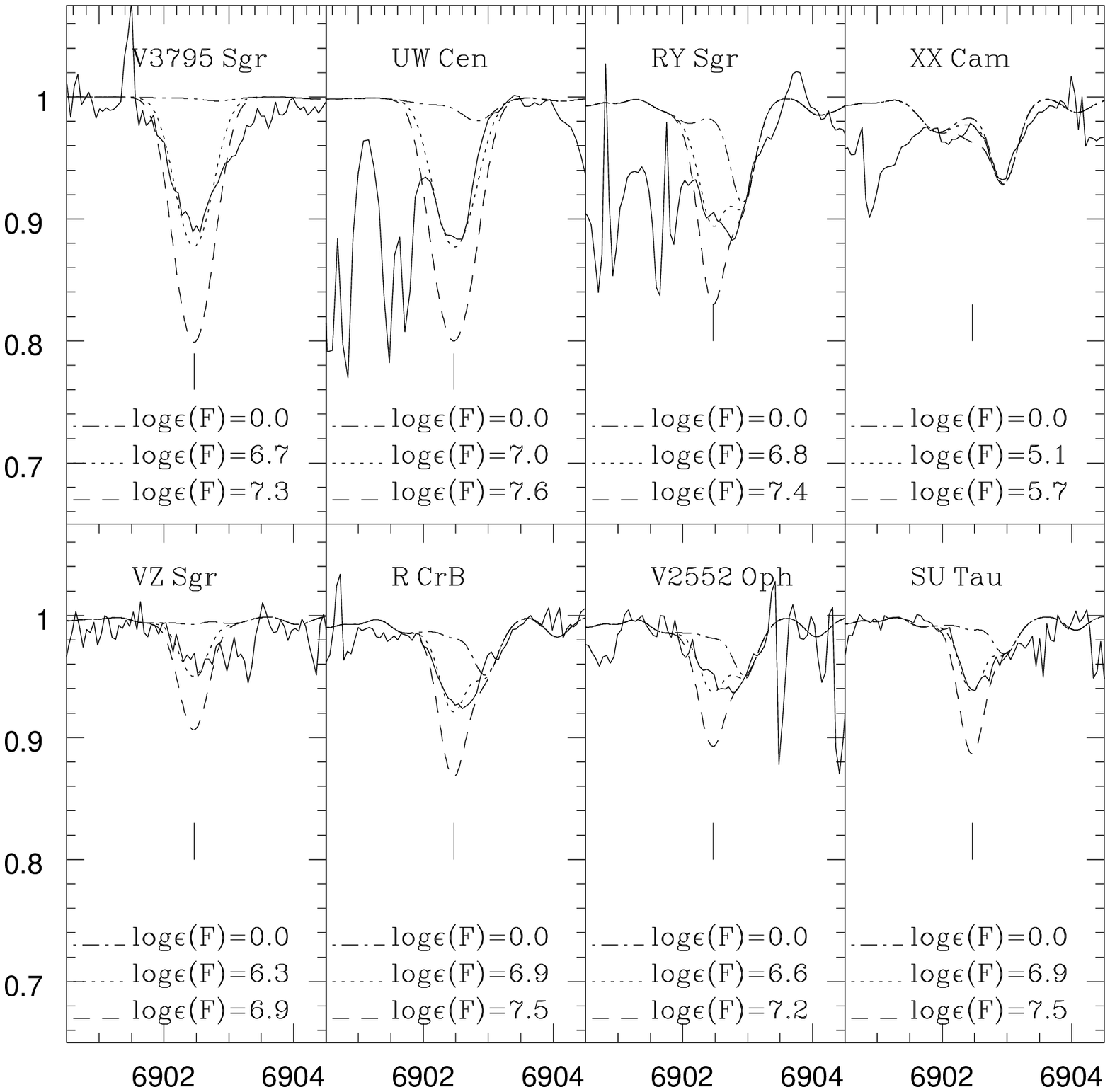}
\caption{Observed F\,{\sc i} 6902.47\AA\ line profiles (solid lines) of
several RCB stars. Synthetic spectra are shown for three fluorine
abundances, as shown on the figure. \label{fig3}}
\end{figure}

\begin{figure}
\epsscale{1.00}
\plotone{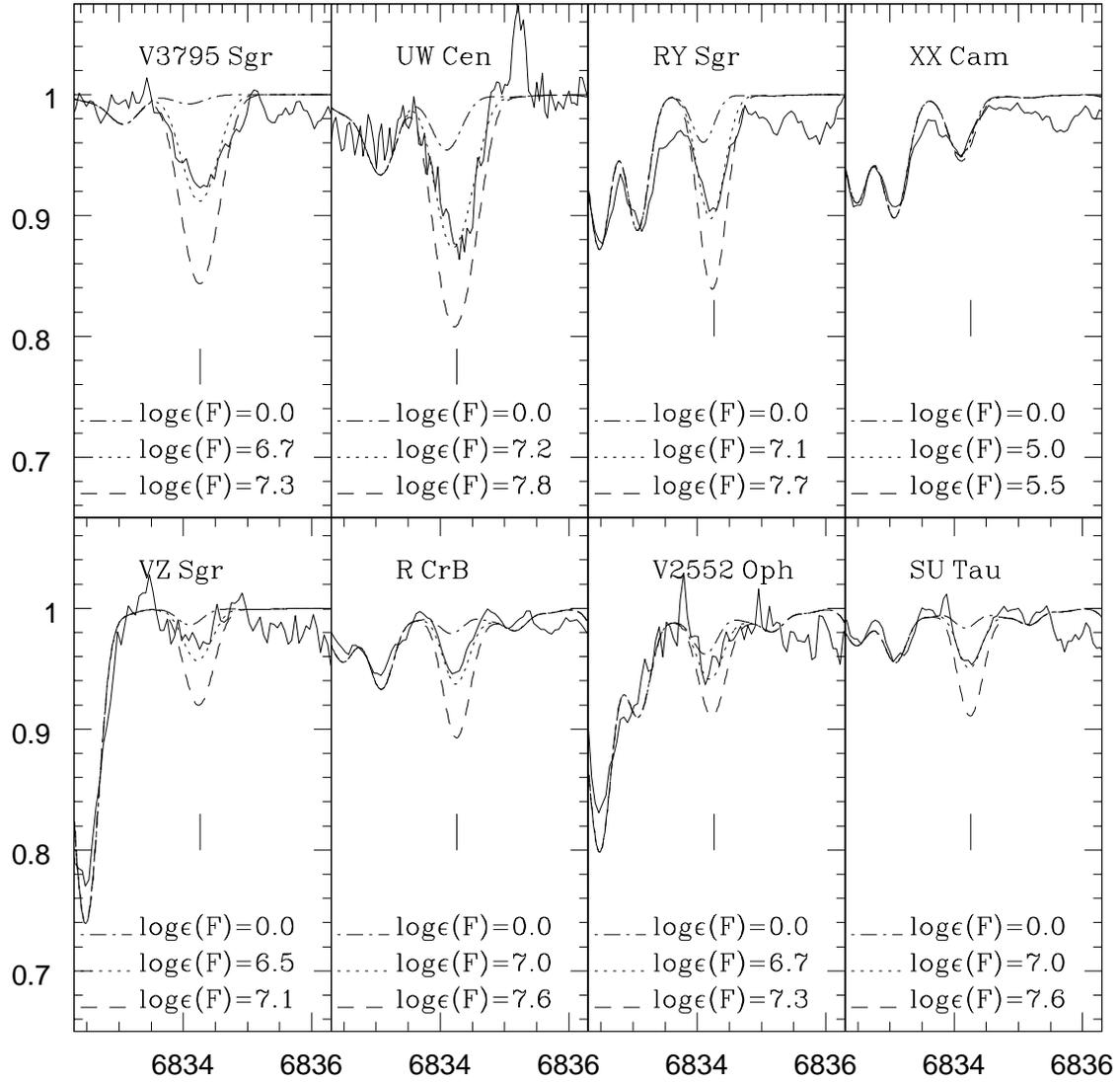}
\caption{Observed F\,{\sc i} 6834.26\AA\ line profiles (solid lines) of
several RCB stars. Synthetic spectra are shown for three fluorine
abundances, as shown on the figure. \label{fig4}}
\end{figure}

\clearpage

The adopted stellar parameters: effective temperature ($T_{\rm eff}$),
surface gravity ($\log g$), and microturbulent velocity ($\xi$),
are from \citet{asp00} for most of the analyzed RCBs. The stellar parameters adopted
for V2552\,Oph, and V\,CrA are from \citet{raolamb2003}, and \citet{raolamb2007},
respectively.
We adopt the Uppsala model atmospheres \citep{asp97a} with a carbon-to-helium abundance ratio C/He
of 1\% for the analyzed RCBs. The C/He ratio is not directly 
determinable for the RCBs. The choice of C/He of 1\% is provided by the measured
C/He of EHe stars, the close relatives of RCBs.
The analyzed EHe stars have a mean C/He of 0.8\% (Pandey et al. 2006 and references 
therein, Pandey \& Reddy 2006), which suggests that the adopted C/He ratio of 1\%
for the Uppsala model atmospheres is a reasonable choice.

%The $gf$-values of the blending
%lines Si\,{\sc i}, C\,{\sc i}, and rest of the lines (see Table 1), are from the
%compilations by R. E. Luck (private communication), NIST database, and Kurucz's database,
%respectively. 
For several of the analyzed RCBs, the best fitting profiles for two of the studied
F\,{\sc i} wavelength regions are illustrated in Figures 3 and 4.
The measured mean fluorine abundances (Table 1) are given as
log $\epsilon(\rm F)$, normalized such that log $\Sigma$$\mu_i \epsilon(i)$ = 12.15
where $\mu_i$ is the atomic weight of element $i$. 

\clearpage

\begin{figure}
\epsscale{1.00}
\plotone{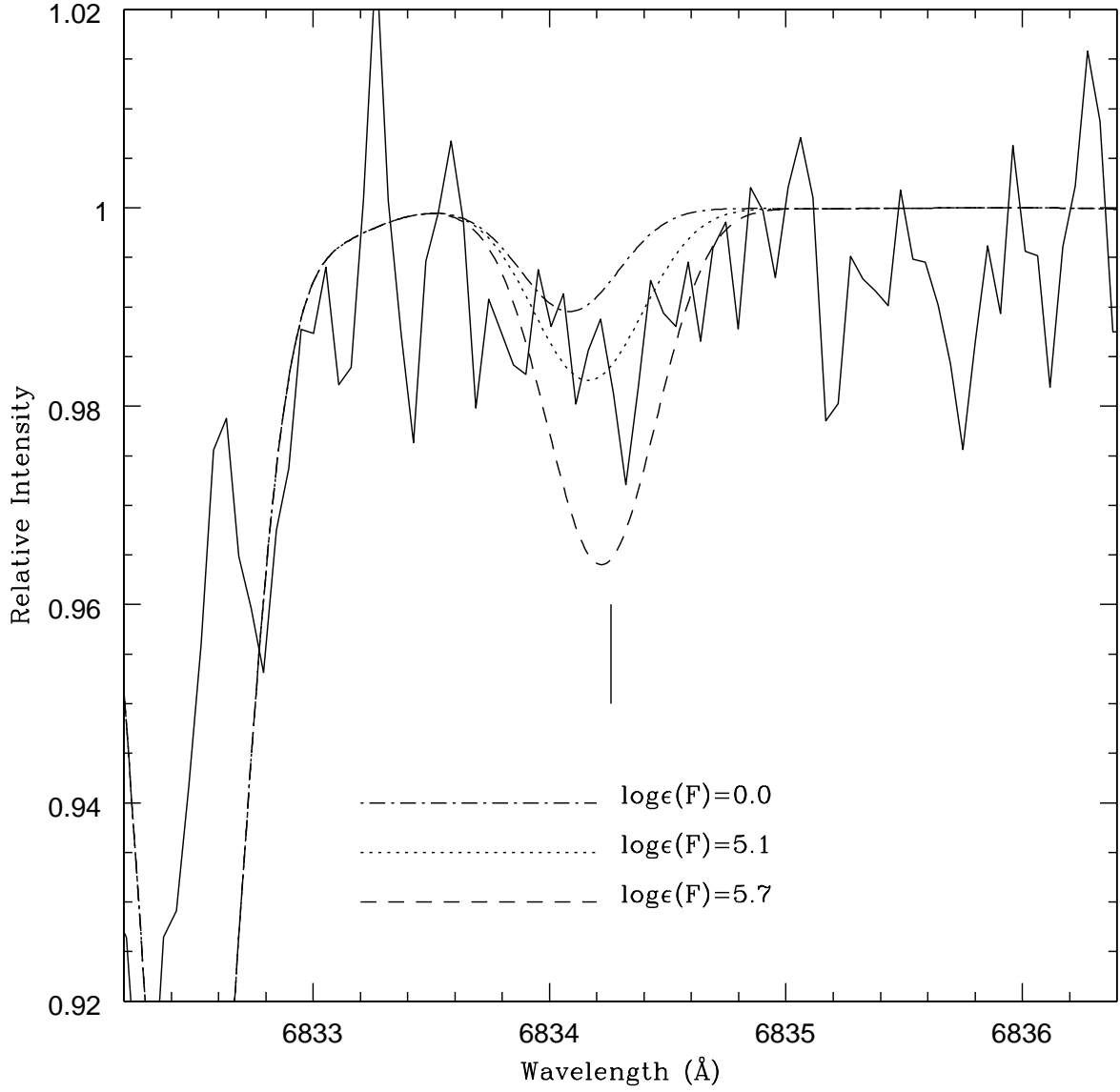}
\caption{Observed F\,{\sc i} 6834.26\AA\ line profile (solid line) of
Sakurai's object. Synthetic spectra are shown for three fluorine
abundances, as shown on the figure. \label{fig5}}
\end{figure}

\clearpage

The errors of the derived  F abundances
given in Table 1 are the line-to-line scatter. The number of lines used is given within
brackets. The F abundances derived for the RCBs cooler or about effective temperature 6500 K 
are uncertain and are marked with a colon in Table 1. However, the F abundances so marked 
are determinations and not  upper limits.
An upper limit to the F abundance for XX\,Cam, V854\,Cen, and the Sakurai's object is obtained 
by comparing the synthetic and the observed profiles of the F\,{\sc i} line at 6834.26\AA\ (Table 1).
The effective error in F abundances due to uncertainty in the
adopted stellar parameters, that are typically: $\Delta$$T_{\rm eff}$ = $\pm$250 K,
$\Delta$$\log g$ = $\pm$0.5 cgs and $\Delta$$\xi$ = $\pm$2 km s$^{-1}$, is about 0.3 dex.
These uncertainties are similar across the sample of analyzed RCBs.
Note that the derived fluorine abundances in LTE represent only the first step defining the
absolute F abundance in these stars; more reliable values should, in principle, come from
full non-LTE calculations.

A spectrum of Sakurai's object (V4334\,Sgr), a final He-shell flash product,
obtained on 5 May 1996 was available for analysis. The nondetection of the 
F\,{\sc i} line at 6834.26\AA\ provides with an upper limit to the F abundance (Figure 5).
Other potential F\,{\sc i} lines in this (then) quite hot star were blended or
in inter-order gaps, e.g., the 6902 \AA\ line is masked by overlying
telluric lines.
Note that the upper limit to the F abundance for the Sakurai's object in Table 1 is for the
model's adopted C/He ratio of 1\%; a model for a C/He ratio of 10\% returns an
upper limit to the F abundance of about 6.4 dex. The stellar parameters
are from \citet{asp97b}.

\section{Discussion}

The analyzed RCBs (excluding stars with upper limits to the F abundances and,
uncertain F abundances) have a mean F abundance of 6.7 which is the same
for the analyzed EHes \citep{pana2006}. Thus, the F abundances are similar across
an effective temperature range from about 6500 K to 14000 K, an indication
that non-LTE effects are possibly small.
 The F abundances of the analyzed RCB and
EHe stars show no obvious trend with their abundances of other elements.
In Figure 6, the F abundance is shown as a function of the Fe abundance for
the RCBs from Table 1 and the EHes from \citet{pana2006}. The F abundances from
RCBs and EHes are in good agreement and independent or almost so of the Fe
abundance which spans a range of nearly two dex. In contrast, the N abundance
is well correlated with the Fe abundance: the N
abundances are consistent with the sum of the initial C, N, and O
abundances suggesting that CNO-cycling has controlled the N
abundances. Among EHes and RCBs, the O
abundance shows a significant scatter at a given Fe abundance. In Figure 7,
the F abundance is shown as a function of the O abundance: the
F abundances are uncorrelated with the O abundance.

\clearpage

\begin{figure}
\epsscale{1.00}
\plotone{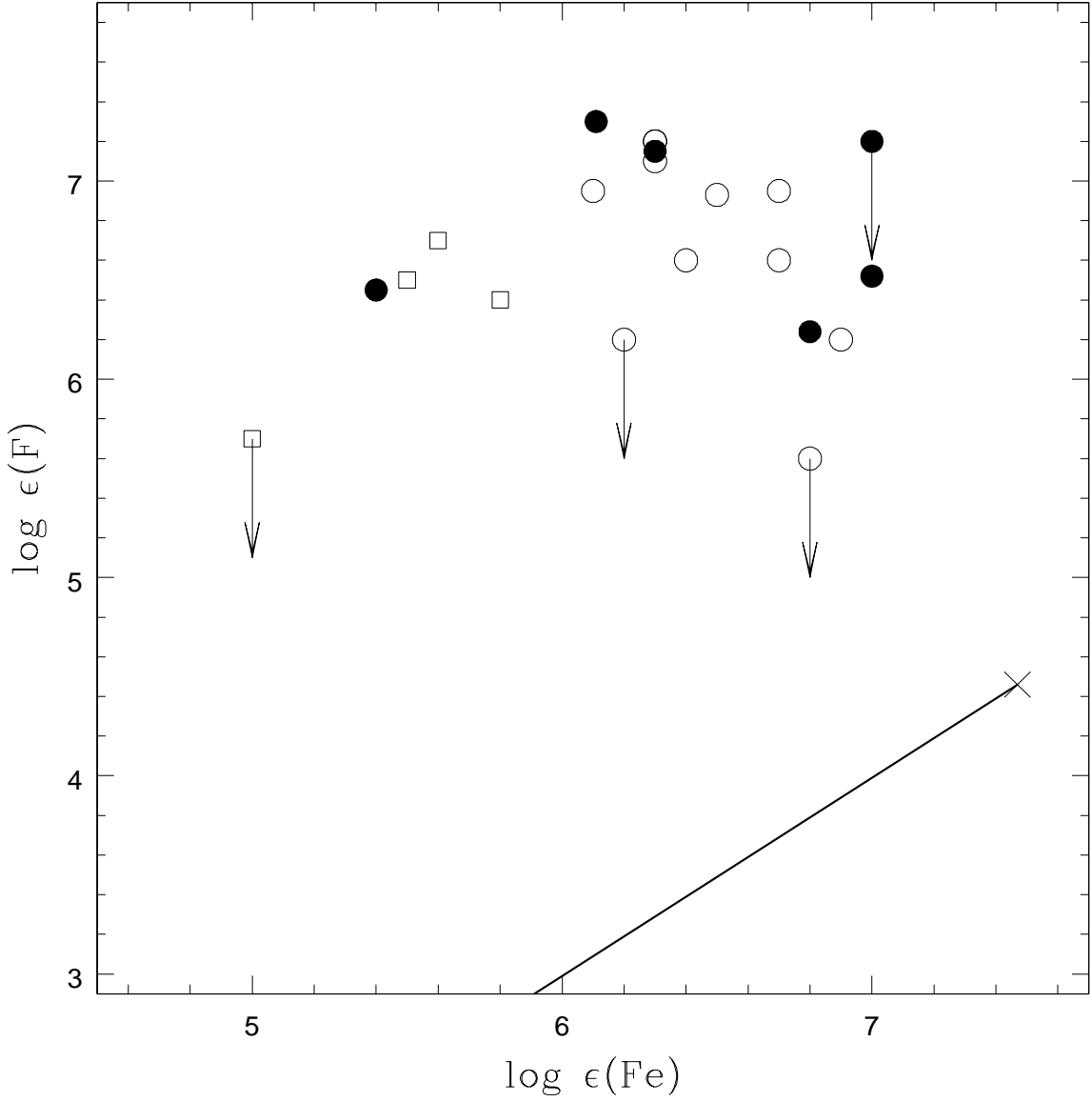}
\caption{The F abundance for majority RCBs (unfilled circles), minority RCBs (unfilled squares), and EHes (filled circles)
as a function of the Fe abundance. The solid line shows the possible
initial F abundances assuming the solar F and Fe abundances (large cross)
and [F/Fe] = 0.
\label{fig6}}
\end{figure}

\begin{figure}
\epsscale{1.00}
\plotone{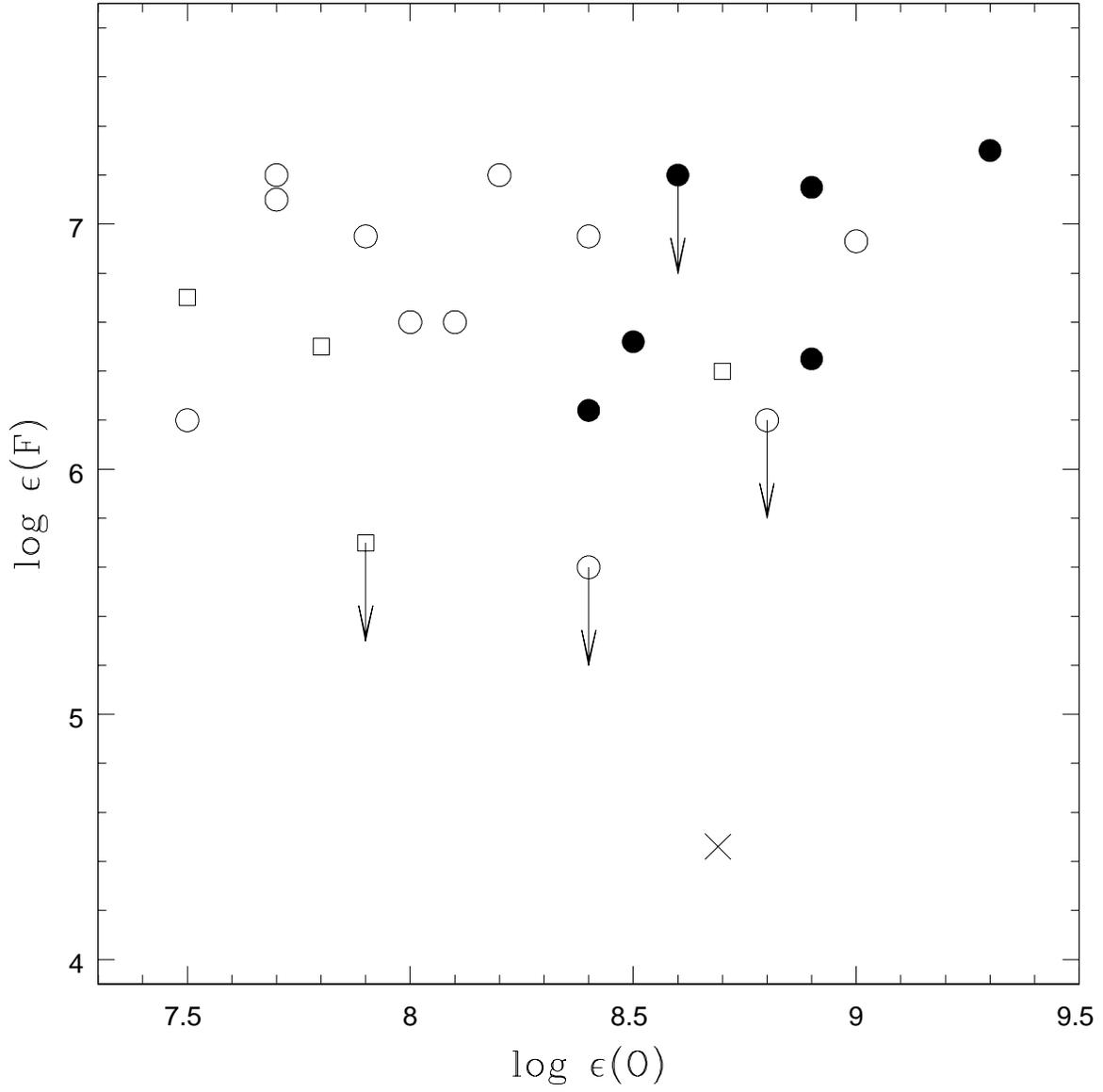}
\caption{The F abundance for majority RCBs (unfilled circles), minority RCBs (unfilled squares), and EHes (filled circles)
as a function of the O abundance.
Solar F and O abundances are represented by the  large cross.
\label{fig7}}
\end{figure}

\clearpage

\subsection{The setting}

Knowledge of fluorine abundance in astronomical objects is limited but,
nonetheless, helpful in interpreting these measurements for 
the RCBs (this paper) and EHes \citep{pana2006}. The solar system
abundance as measured from
meteorites is $\log\epsilon$(F) = 4.46$\pm0.06$ \citep{lod2003,asp2005}
with rather imprecise confirmation from the detection of HF lines in
sunspots. Measurements of HF lines in K and M Galactic
giants \citep{jori92,cunha2003} give a mean F abundance of $\log\epsilon$(F) = 4.62$\pm0.12$
for a sample with a mean Fe abundance ([Fe/H]$=-0.10\pm0.16$) not
significantly different from solar. The trend of F with Fe is quite uncertain.
The sole  Galactic star with a (slightly) sub-solar Fe abundance is
Arcturus with [Fe/H]=$-$0.6 and a F abundance ($\log\epsilon$(F) = $4.10$ -
Cunha et al. 2003) suggesting that [F/Fe] $\sim 0$ but observations of
additional metal-poor giants are required to confirm this hint.

The F vs Fe trend is one important datum  for inferring the RCBs initial
F abundance from their present Fe abundances.
The mean Fe abundance
from \citet{asp00} and \citet{raolamb2003} for 14 majority
RCBs is $\log\epsilon$(Fe) = $6.5$ with a spread of 0.8 dex; three
minority RCBs (V\,CrA, VZ\,Sgr, and V3795\,Sgr) of very low Fe
abundance are not included here, however, these three minority RCBs
are not different in F abundances when compared with the majority RCBs. 
To this mean Fe abundance must
be attached the qualification that the analyses assume a C/He ratio of
1\% by number but the analyses do not return the input C abundance but a
value about 0.6 dex lower, a discrepancy termed `the carbon problem' by
Asplund et al.  Analyses of EHe stars (Pandey et al. 2006 and references
therein) which are immune to the carbon problem give a mean Fe
abundance $\log\epsilon$(Fe) = $6.7$ and a spread of 1.1 dex with two
minority EHes excluded. The similar Fe abundances
of EHes and RCBs suggests that the carbon problem is not
adversely affecting the determination of the Fe abundance.
The mean C/He ratio found for the EHes is slightly less than the assumed
C/He = 1\% for the RCBs. Adoption of the C/He ratio of the EHes would
reduce the mean Fe abundance for the RCBs by about 0.15 dex.
The assumption is made that that the Fe abundance is not
affected by the stars' evolutionary history.
It is striking that the EHes and RCBs are quite Fe-poor
relative  to the Sun. This might occasion concerns that the initial iron
abundance has been reduced in the process of evolution to a H-poor
star. In this connection, we note that abundance ratios [X/Fe],
especially those of the $\alpha$-elements  -- Mg, Si, S and Ca --
are those expected of metal-poor stars and not those of solar
abundance stars \citep{asp00,pan2006}. In the
following discussion, we assume that the present abundances of iron
are the initial values.

Although, as noted above, the observational
 data are sparse in the extreme, it seems
safe to assume that [F/Fe] $\leq 0$ for metal-poor ([Fe/H]$<0$) stars.
Galactic chemical evolution models confirm this expectation \citep{renda2004}.
In estimating the F enrichment for EHes and
RCBs, we assume [F/Fe]=0 and the solar F abundance
$\log\epsilon$(F)$ = 4.46$. Thus, a representative RCB with 
$\log\epsilon$(Fe)$ = 6.5$ or [Fe/H]= $-1.0$ is assumed to have
an initial F abundance of $\log\epsilon$(F)$ \leq 3.5$.
More interestingly, the F overabundances are extremely large:
enhancements of about 800, 2500, and 8000 at $\log\epsilon$(Fe)$ = 7.5$, 6.5, and 5.5,
respectively. The fact that the overabundances are very large is qualitatively
true irrespective of all plausible values for [F/Fe] as a function of [Fe/H].

\subsection{The DD Scenario and Fluorine}

In the `cold' (i.e., no nucleosynthesis during the merger) version of the
DD scenario, the He white dwarf is stripped and accreted by the
C-O white dwarf. During the active phase of the merger, the accreted
material is mixed with the outer layer of the C-O white dwarf
which is the residue of the He-intershell of an AGB star for
which the C-O white dwarf was the electron degenerate core. Mixing
may also  penetrate the outer layers of the C-O core. In the `hot' DD scenario, nucleosynthesis
occurs during accretion. Following both cold and hot mergers, the hot luminous
H-poor star evolves to the red before returning to the blue and descending the
white dwarf cooling track. Along this track, He-shell flashes are likely to
occur and dredge-ups are likely to alter the surface chemical evolution.
These alterations have not been extensively studied theoretically.
Limited predictions for a slow rate of accretion have been reported
by \citet{saio02}. 

Given the principal features of the expected compositions of the
He white dwarf and the He-shell around the C-O white dwarf, one
simple recipe that provides an atmosphere with a RCB- and EHe-like
composition is as follows:
a 0.3$M_\odot$ He white dwarf is mixed with 0.03$M_\odot$ of
He-intershell material. This cold recipe does not consider nucleosynthesis
to occur during or following the merger.
 The relative masses of He-intershell and He white dwarf
determine the final C/He; the He white dwarf is
rich in He but  highly deficient
in C and O and the He-intershell is relatively rich in C and O. The N abundance
is provided primarily by the He white dwarf.
  A H-rich surface layer on
the He and/or the C-O white dwarf provides the residual H for the RCB and EHe.
For discussion of such a recipe and how it may match observed
compositions of RCBs and EHes, see \citet{saio02} and \citet{pan2006}.

For elements greatly enriched in the He-intershell, their
abundance in the RCB and EHe is diluted by the
material from the He white dwarf. A dilution by a factor of order
ten is necesssary in light of the expected abundances of C and He in the
He-intershell. In the case of F, we may appeal to theoretical
predictions of F abundances in the He-intershell of AGB stars
\citep{lug2004} and to observations of F in PG1159 stars.
PG1159 stars are H-deficient very hot ($T_{\rm eff} = 75,000 - 200,000$ K)
post-AGB stars whose surfaces `offer the unique possibility of
studying intershell abundances {\it directly}' \citep{werner2006}.
The PG1159 stars are considered to be products of the FF scenario
and to involve a `very late thermal pulse' (VLTP) (see Werner \& Herwig [2006]
for the difference between a very late thermal pulse and a late thermal
pulse). The C/He ratio of PG1159 stars is too high for EHe and RCB stars to
be identified with the FF scenario. In the DD scenario, the high C/He ratio
is diluted to a lower ratio when the FF-like layer of the C-O white dwarf
is mixed with the He white dwarf.

Fluorine in PG1159 stars shows a range of abundances \citep{werner2005,werner2006}
from solar to 250 times solar.
As Werner \& Herwig discuss, this
range is not out of line with theoretical predictions for the He-intershell. In addition,
the theoretical predictions account fairly well for the observed enrichments
of F in (H normal) giant stars known to be affected by the third dredge-up that
brings material from the He-intershell into the envelope \citep{jori92}.
But in the cold DD scenario, the He-intershell material
is diluted, according to the canonical recipe, by a factor of about ten and,
then, overabundances of up to 25 times solar for the
RCBs and EHes are predicted. The observed overabundances of F range upward of 1000 times.
Unless the  VLTP experienced by the PG1159 stars drastically
reduced the F abundance of the He-intershell or the nucleosynthesis
experienced by the He-intershell of the C-O white dwarf differs
markedly from that experienced by the PG1159 stars, it would appear
that the cold DD scenario can  not account for the observed F
abundances of RCBs and EHes.

The upper limit for the F abundance of Sakurai's object is consistent
with but does not demand that ingestion of H and the final
He-shell flash leads to a complete conversion of 
$^{14}$N to $^{22}$Ne. The temperatures are so 
high that $^{18}$O, which is the seed for $^{19}$F, is destroyed.
Note that the infrared spectra of Sakurai's object obtained
during 1997$-$1998 when it had strong CO overtone bands showed no
evidence for $^{12}$C$^{18}$O \citep{geballe2002}.
Therefore, Sakurai's object and the other FF stars, when compared
with most of the RCB and HdC stars, probably have different origins.

A `hot' DD scenario is fuelled by a rapid deposition of accreted
He-rich material on the surface of the C-O white dwarf.
Observational evidence that suggests
a hot DD scenario is provided by Clayton et al.'s (2005, 2007) discovery
of huge overabundances of $^{18}$O in HdC and some cool RCBs through
medium-resolution spectroscopy of the first-overtone vibration-rotation
bands of the CO molecule near 2.3 microns. High-resolution
spectroscopy confirms that $^{16}$O/$^{18}$O ratios are low:
\citet{garc2007} find $^{16}$O/$^{18}$O $=0.3$
to 0.5 for the three HdCs with CO bands and 16 for the cool RCB S\,Aps.

\citet{clay2007} show via qualitative arguments about temperature, density, and
timescale that the merger leads to temperatures at the base of the
accreted material approaching 200 million K with conditions 
ripe for nucleosynthesis lasting several years.
Their calculations with a reaction network show that the
plentiful supply of $^{14}$N in the He white dwarf may be converted
 to $^{18}$O by $\alpha$-capture.
Without this conversion, the DD scenario cannot account for the $^{18}$O
overabundances; the expected
$^{18}$O abundance in the He white dwarf and in  the He-intershell
of the C-O white dwarf are very low. The present measurements of
$^{18}$O in HdC and RCB stars indicate that no more than a minor fraction
of the $^{14}$N is converted to $^{18}$O.
(`No overproduction of $^{18}$O is
expected in the FF scenario' state Clayton et al. [2007].)

The possibilities for nucleosynthesis in the hot DD scenario are
more extensive if the He white dwarf has a H-rich outer layer.
Since many HdC, RCB, and EHe stars have retained some H, H
was likely a participant in the nucleosynthesis. \citet{clay2007}
note that the presence of protons in $^{18}$O-rich
gas promotes the synthesis of
$^{19}$F. The reaction  $^{18}$O$(p,\gamma)^{19}$F provides
$^{19}$F directly.
% with an additional channel involving
%$^{18}$O(p,$\alpha$)$^{15}$N and $^{15}$N($\alpha$,$\gamma$)$^{19}$F.
The $^{18}$O is converted about 150 times faster by proton-captures to $^{12}$C:
$^{18}$O(p,$\alpha$)$^{15}$N(p,$\alpha$)$^{12}$C. 
Clayton et al.'s (2007, their Figure 6)  sample single-zone
 calculation for a mixture with H added
and for  an arbitrary temperature history and initial composition
showing effects of some He-burning
 shows that
the $^{19}$F abundance increased by about a dex at the cessation of
nucleosynthesis from an initial value about a factor of three below
solar to a factor of two above solar. At its peak, F was briefly
about 100 times above its solar abundance before $^{19}$F($\alpha$,p)$^{22}$Ne
took its toll. 
 Clearly, although $^{19}$F synthesis is demonstrated
by these calculations, a remaining challenge is to show that the parameter
space covered by the hot DD scenario includes the possibility of
robustly increasing the F abundance not by a factor of 100 but
 to the observed levels of 1000 to 8000
times over solar across the metallicity range [Fe/H] of $-0.5$ to $-2.0$.

Fluorine abundances appear to demand that H be a key  ingredient in
the hot DD scenario.  This is suggested by the
  F/N ratios. The N abundances of the
EHe stars are equal to the sum of the initial C, N, and O abundances,
as expected for He-rich material exposed to CNO-cycling \citep{pan2006}.
The RCBs give a similar result.  The observed N/F ratios are about
80, 60, and 5 at [Fe/H] $= -0.5$, $-1.0$, and $-2.0$, respectively.
Given that about 150 times more $^{18}$O is diverted to $^{12}$C than is
converted to $^{19}$F, the observed N/F abundances imply that 
total conversion of the  $^{14}$N from the He white dwarf to $^{18}$O
cannot provide the observed $^{19}$F abundances and this total
conversion certainly cannot account for the fact that the $^{14}$N
across the sample of EHe and RCB stars
is consistently close to that
expected from CNO-cycling for the He white dwarf.
The conclusion  seems to be that hot accretion must involve
addition of H to the extent that N  and $^{18}$O are  regenerated by H-burning.
This and other
nucleosynthetic consequences are illustrated by
Clayton et al. (2007, their Figure 6). Their
calculation if representative would explain the star-to-star
variation in O abundances but the fact that the observed abundances
are close to the sum of the initial C, N, and O abundances has then to
be regarded as fortuitous. Obviously, challenges remain in accounting
for the F abundances. One challenge arises from incorporating Clayton et al.'s
single-zone calculations into a full  model of the accreting star and
its subsequent evolution  
with  attendant shell flashes and dredge-ups.  

Another challenge involving nucleosynthesis concerns the abundance of
Li. Lithium is present in four of the analyzed 15 majority RCBs \citep{raolamb94}
and in one of the five HdCs \citep{raolamb96} with
an appreciable abundance  $\log\epsilon$(Li) $\sim 3$. The only
conceivable reservoir of Li in the two white dwarfs ahead of their
merger is the H-rich layer around the He white dwarf but the mass of this
is too small to result in the above Li abundances, even if the H-rich
layer had retained its initial Li abundance. Lithium synthesis is
presumed to involve $^3$He and the reactions
$^3$He($\alpha$,$\gamma$)$^7$Be(e,$\nu$)$^7$Li.
The $^3$He may be present in the H-rich layer as
the product of the
first steps of the $pp$-chain and as an initial ingredient of the star.
In a hot DD scenario in which $^{14}$N is converted to $^{18}$O, it
seems inconceivable that $^7$Be or $^7$Li can avoid thorough
destruction by $\alpha$-capture; enrichment of Li in parallel with 
enrichment of $^{18}$O and F seems impossible in nucleosynthesis 
accompanying the DD scenario \citep{clay2007}.
One supposes that Li may be synthesized in `warm' DD scenarios in which
mild nucleosynthesis occurs with temperatures too low for $^{18}$O
production. Lithium synthesis is possible in the FF scenario.
Most irritatingly, the
HdC and RCB stars for which $^{18}$O has been detected are not those
in which Li is present; note that the possibility of $^{18}$O
enrichment in the Li-rich HdC star HD\,148839 is not discarded
by \citet{garc2007}.
However, the Li-rich RCB stars are not
distinguishable from other RCBs and all have the He/C/N/O ratios
of a DD scenario and incompatible with a FF scenario.
It is intriguing that the Li-rich stars (UW\,Cen, R\,CrB, and SU\,Tau) are among
the stars with the highest F abundances. The detection of high F in both
minority and majority RCBs is also very interesting since $^{18}$O
observations do not include any of these stars as these are too hot.
This adds the minority RCBs as well as the Li-rich RCB stars to the
stars that might have large amounts of $^{18}$O.

\section{Concluding remarks}

Lithium and fluorine are not the sole or even the principal
concerns for the DD scenario.
One wonders why the C/He $\sim 0.01$ ratio, as measured for the EHes, spans
a narrow range.  The He is contributed  primarily by the He white
dwarf and the C by the He-intershell of the C-O white dwarf. What sets
the mass ratio of these two regions rather precisely?

The evolutionary connections between the H-deficient
stars remains unproven but with obvious potential
links. Both possible sequences --
 HdC $\rightarrow$ RCB $\rightarrow$ EHe and
EHe $\rightarrow$ RCB $\rightarrow$ HdC are possible
because the DD scenario leads to a very hot star
evolving to lower temperatures before the evolutionary track reverses
direction.
An extension EHe $\rightarrow$ sdO $\rightarrow$ O(He)
to higher temperatures and higher surface gravities  seems
plausible \citep{rauch2006,werner2006}.
The stars in this sequence are all candidates for an origin via the
DD scenario, i.e., they have the low C/He ratio and not the high
ratio characteristic of the the FF scenario; a similar
sequence for the FF scenario exists  linking objects like FG\,Sge and
V4334\,Sgr (Sakurai's object) through the [WC] to the PG1159 stars.

Continued study of the compositions of stars along these
sequences will serve to sharpen understanding of the evolutionary
links and the entry points  for stars into the sequence.
New data such as that for F and $^{18}$O stimulate continued
study. The F abundances of  RCB  and EHe stars are quite similar in
value and spread and thus support the presumed evolution from
RCB to EHe. On the other hand, there is a hint that the
$^{16}$O/$^{18}$O ratios of the HdC and RCB stars may be
different with the ratio being systematically higher in the
(cool) RCBs than in the three HdCs. Is the difference merely
a distraction caused by small number statistics or is it
symptomatic of a change introduced as a HdC evolves to a cool RCB or
does it reflect that the DD scenario can place stars on the evolutionary
sequence at different starting points with different compositions?

We thank the referee Geoff Clayton for the nice report. 
GP thanks the VBT staff for their assistance.
This research has been supported in part by the Robert A. Welch Foundation
of Houston, Texas.


\clearpage

\begin{thebibliography}{}
\bibitem[Asplund et al.(1997a)]{asp97a} Asplund, M., Gustafsson, B., Kiselman, D.,
\& Eriksson, K. 1997a, \aap, 318, 521
\bibitem[Asplund et al.(1997b)]{asp97b} Asplund, M., Gustafsson, B., Lambert, D. L.,
\& Rao, N. K. 1997b, \aap, 321, L17
\bibitem[Asplund et al.(2000)]{asp00} Asplund, M., Gustafsson, B., Lambert, D. L.,
  \& Rao, N. K. 2000, \aap, 353, 287
\bibitem[Asplund et al.(2005)]{asp2005} Asplund, M., Grevesse, N., 
\& Sauval, A. J. 2005, ASP Conf. Ser. 336, 25
\bibitem[Clayton(1996)]{clay96} Clayton, G. C. 1996, \pasp, 108, 225 
\bibitem[Clayton et al.(2005)]{clay2005} Clayton, G. C., Herwig, F., Geballe, T. R., Asplund, M.,
Tenenbaum, E. D., Engelbracht, C. W., \& Gordon, K. D. 2005, \apjl, 623, L141 
\bibitem[Clayton et al.(2007)]{clay2007} Clayton, G. C., Geballe, T. R., Herwig, F., Fryer, C.,
\& Asplund, M. 2007 \apj, 662, 1220
\bibitem[Cunha et al.(2003)]{cunha2003} Cunha, K., Smith, V. V., Lambert, D. L.,
\& Hinkle, K. H. 2003, \aj, 126, 1305
\bibitem[Geballe et al.(2002)]{geballe2002} Geballe, T. R., Evans, A., Smalley, B., 
Tyne, V. H., Eyres, S. P. S. 2002, Ap\&SS, 279, 39
\bibitem[Guerrero et al.(2004)]{gue04} Guerrero, J.,
Garc\'{\i}a-Berro, E., \& Isern, J. 2004, \aap, 413, 257
\bibitem[Garc\'{\i}a-Hern\'{a}ndez et al.(2007)]{garc2007} Garc\'{\i}a-Hern\'{a}ndez, D.A., 
Hinkle, K.H., Lambert, D.L., \& Eriksson, K. 2007, \apj, submitted
\bibitem[Jorissen et al.(1992)]{jori92} Jorissen, A., Smith, V. V., \&
Lambert, D. L. 1992, \aap, 261, 164
\bibitem[Lambert \& Rao(1994)]{raolamb94} Lambert, D. L., \& Rao, N. K. 1994, JAA, 15, 47
\bibitem[Lodders(2003)]{lod2003} Lodders, K. 2003, \apj, 591, 1220
\bibitem[Lugaro et al.(2004)]{lug2004} Lugaro, M., Ugalde, C., \&
Karakas, A., I. 2004, \apjl, 615, L934
\bibitem[Moore(1965)]{moore65} Moore, Ch. E. 1965, Selected Tables of Atomic Spectra,
NSRDS--NBS 3, Section 2, Washington
\bibitem[Moore(1967)]{moore67} Moore, Ch. E. 1967, Selected Tables of Atomic Spectra, 
NSRDS--NBS 3, Section 1, Washington
\bibitem[Moore(1972)]{moor72} Moore, Ch. E. 1972, A Multiplet Table of Astrophysical
Interest, NSRDS--NBS, Washington
\bibitem[Moore(1993)]{moor93} Moore, Ch. E. 1993, Tables of Spectra of Hydrogen, Carbon, 
Nitrogen, and Oxygen Atoms and Ions, Editor: Jean W. Gallagher, CRC Series in Evaluated Data
in Atomic Physics, CRC press 
\bibitem[Musielok et al.(1999)]{musie99} Musielok, J., Pawelec, E., Griesmann, U., \&
Wiese, W. L. 1999, \pra, 60, 947
\bibitem[Nave et al.(1994)]{nave94} Nave, G., Johansson, S., Learner, R. C. M., 
Thorne, A. P., \& Brault, J. W. 1994, \apjs, 94, 221
\bibitem[Pandey et al.(2001)]{pan2001} Pandey, G., Rao, N. K., Lambert, D. L.,
Jeffery, C. S., \& Asplund, M. 2001, \mnras, 324, 937
\bibitem[Pandey et al.(2004a)]{pan2004a} Pandey, G., Lambert, D. L., Rao, N. K.,
\& Jeffery, C. S. 2004a, \apjl, 602, L113
\bibitem[Pandey et al.(2004b)]{pan2004b} Pandey, G., Lambert, D. L., Rao, N. K.,
Gustafsson, B., Ryde, N., \& Yong, D. 2004b, \mnras, 353, 143
\bibitem[Pandey et al.(2006)]{pan2006} Pandey, G., Lambert, D. L.,
Jeffery, C. S., \& Rao, N. K. 2006, \apj, 638, 454
\bibitem[Pandey \& Reddy(2006)]{panred2006} Pandey, G., \& Reddy, B. E. 2006,
\mnras, 369, 1677
\bibitem[Pandey(2006)]{pana2006} Pandey, G. 2006, \apjl, 648, L143
\bibitem[Rao \& Lambert(1996)]{raolamb96} Rao, N. K., \& Lambert, D. L. 1996,
ASP Conf. Ser. 96, 43
\bibitem[Rao \& Lambert(2000)]{raolamb2000} Rao, N. K., \& Lambert, D. L. 2000,
\mnras, 313, L33
\bibitem[Rao \& Lambert(2003)]{raolamb2003} Rao, N. K., \& Lambert, D. L. 2003,
\pasp, 115, 1304
\bibitem[Rao et al.(2004)]{rao04} Rao, N. K., Sriram, S., Gabriel, F., Prasad,
B. R., Samson, J. P. A., Jayakumar, K., Srinivasan, R., Mahesh, P. K.,
\& Giridhar, S. 2004, Asian Journal of Physics, 13, 267
\bibitem[Rao(2005)]{rao2005} Rao, N. K. 2005, ASP Conf. Ser. 336, 185
\bibitem[Rao et al.(2005)]{rao05b} Rao, N. K., Sriram, S., Jayakumar, K.,
\& Gabriel, F. 2005, JAA, 26, 331
\bibitem[Rao et al.(2006)]{raolambsh2006} Rao, N. K., Lambert, D. L., 
\& Shetrone, M. D. 2006, \mnras, 370, 941 
\bibitem[Rao \& Lambert(2007)]{raolamb2007} Rao, N. K., \& Lambert, D. L. 2007,
\mnras, in press
\bibitem[Rauch et al.(2006)]{rauch2006} Rauch, T., Reiff, E., Werner, K., Herwig, F.,
Koesterke, L., \& Kruk, J. W. 2006, ASP Conf. Ser. 348, 194
\bibitem[Renda et al.(2004)]{renda2004} Renda, A., et al. 2004, \mnras, 354, 575
\bibitem[Saio \& Jeffery(2002)]{saio02} Saio, H., \& Jeffery, C. S. 2002, \mnras, 333, 121
\bibitem[Tenenbaum et al.(2005)]{tenen2005} Tenenbaum, E. D., Clayton, G. C., 
Asplund, M., Engelbracht, C. W., Gordon, K. D., Hanson, M. M., Rudy, R. J., 
Lynch, D. K., Mazuk, S., Venturini, C. C., \& Puetter, R. C. 2005, \aj, 130, 256
\bibitem[Tull et al.(1995)]{tull95} Tull, R. G., MacQueen P. J., Sneden, C., \&
Lambert, D. L. 1995, \pasp, 107, 251
\bibitem[Werner et al.(2005)]{werner2005} Werner, K., Rauch, T., \&
Kruk, J. W. 2005, \aap, 433, 641
\bibitem[Werner \& Herwig(2006)]{werner2006} Werner, K., \& Herwig, F. 2006,
\pasp, 118, 183
\bibitem[Zaniewski et al.(2005)]{zanie2005} Zaniewski, A., Clayton, G. C., Welch, D. L., 
Gordon, K. D., Minniti, D., \& Cook, K. H. 2005, \aj, 130, 2293


\end{thebibliography}
\end{document}